# Learning about End-User Development for Smart Homes by "Eating Our Own Dog Food"

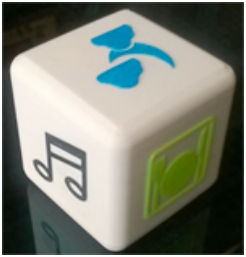

**The DomiCube.** The DomiCube is a home-made device designed by 5 retired seniors as the result of a 3 hour focus group. It contains an accelerometer and a gyroscope, and is Bluetooth enabled. It sends events when its state changes (e.g., new orientation, top face, and battery level). The DomiCube was built in the Creativity Lab of the EquipEx AmiQual4Home, ANR-11-EQPX-00.

**Joëlle Coutaz**
University of Grenoble.
Laboratory of Informatics of
Grenoble (LIG), France
joelle.coutaz@imag.fr

**James L. Crowley**
University of Grenoble,
Laboratory of Informatics of
Grenoble (LIG), INRIA Grenoble
Rhônes-Alpes Research Center,
France
james.crowley@inria.fr

The system can be demonstrated at the workshop provided that Internet access is available.

## Abstract
SPOK is an End-User Development Environment that permits people to monitor, control, and configure smart home services and devices. SPOK has been deployed for more than 4 months in the homes of 5 project team members for testing and refinement, prior to longitudinal experiments in the homes of families not involved in the project. This article reports on the lessons learned in this initial deployment.

## Author Keywords
End-user programming; end-user development; connected home; smart home; ubiquitous computing.

## ACM Classification Keywords
D.2.6. Programming Environments: Interactive environments. H.5.2. User Interfaces.

## Introduction
The "Do-It-Yourself" approach to configuring and controlling domestic technology has become increasingly popular. End-User Development Environments (EUDE) have been developed to support this approach. While the Scratch-based programming language [3] used in the ZipaBox and the rule-based IFTTT propose attractive graphical syntax and stylistics,

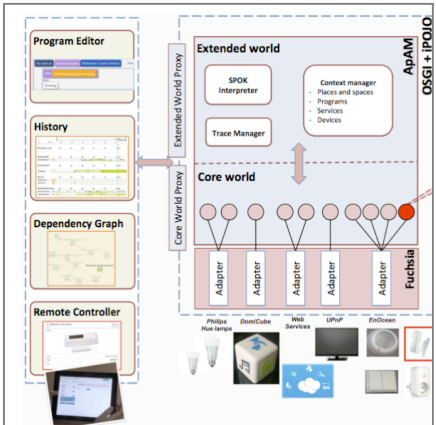

**Global architecture.** The SPOK **server** runs on top of OSGi/iPOJO augmented with ApAM (right side of the figure). It is structured as two levels of abstraction: (a) The Core World that abstracts away devices such as the Philips Hue lights, home-made DomiCube, physical sensors and actuators supported by a diversity of communication protocols (UPnP, MQTT, EnOcean, BlueTooth), and cloud services (Yahoo! Weather forecast, Google Calendar, gmail); (b) The Extended world whose functional coverage depends on the application domain (e.g., SPOK EUDE). The server runs either on a MiniPC, Raspberry Pi, or a Google Nexus Tablet. **Clients** (left side of the figure) run a web browser to access SPOK services.

these environments provide limited debugging aids for non-specialists. At the same time, seminal work from academic research such as Jigsaw [2], CAMP [4], and iCAP [1], has not gone beyond proof of concept. Thus, over the last two years, we have developed SPOK (Simple PrOgramming Kit), an EUDE for smart homes. Our goal is to provide a robust, extensible and flexible system that can be used effectively in the home, and that can be evaluated through longitudinal experiments in real life conditions.

In the next section, we explain our technical choice for the baseline middleware for SPOK followed by a presentation of the principal features. In the final section, we summarize the key lessons learned from the initial deployment of SPOK in our own homes.

## Beyond Technical Proof-of-Concept

Implementing a EUDE for real world smart homes requires choosing the "appropriate" run time infrastructure from a jungle of middleware. From our experience with the development of an earlier environment KISS [5], "appropriate", in a research context, means: (1) license free and robust, (2) support for dynamic discovery, (re)composition and deployment without human intervention, and (3) low entry cost for developers.

OSGi satisfies the two first requirements but is too low-level for non-system developers. OpenHAB [14], which has been chosen by Eclipse as the Eclipse SmartHome, does not, in its current form, support dynamicity. HomeOS is by construction a .net environment, therefore not compatible with iOS and Android platforms. In the absence of a *de facto* standard middleware for smart homes, we have used ApAM.

ApAM (Application Abstract Machine) is a component-oriented middleware that extends OSGi/iPOJO in two ways: (1) developers describe an application by intention using a dedicated language as opposed to explicitly specifying composition of components and bindings at design time; (2) from the abstract description of the application architecture, a concrete architecture is computed and incrementally updated by resolving the dependencies between the components currently available in the execution environment.

Due to the incremental and dynamic (just-in-time) construction and maintenance made possible by ApAM, SPOK is resilient to the opportunistic installation and disappearance of devices and services. The sidebar shows the overall description of the global architecture of SPOK.

## The Description of SPOK

SPOK provides the end-user with the following services: (1) A syntax-oriented program editor that enforces the construction of syntactically-correct programs (see sidebar on next page). (2) A program interpreter and a clock simulator to test program execution in "simulated time". (3) Debugging aids to support the detection and correction of programming errors or system malfunctions along with a Trace Manager. (4) A dashboard to remotely control devices and programs in a centralized and uniform manner.

Compared to the state-of-the art, the key features of SPOK are thee-fold: Expressive power of the SPOK language along with a pseudo-natural concrete syntax, dynamic adaptation to the arrival/departure of devices and services, and debugging aids.

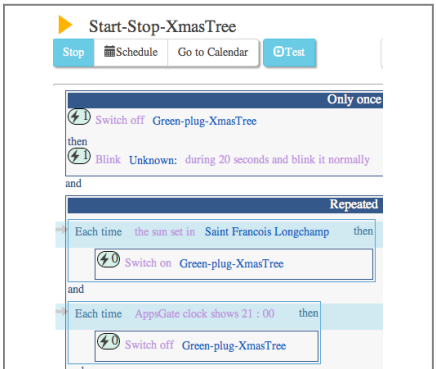

**Start-Stop-XmasTree is being executed.**

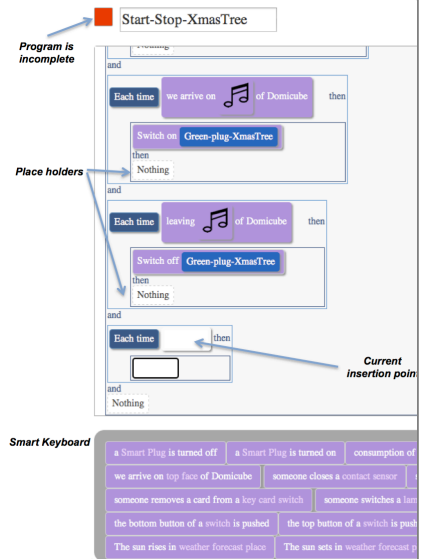

**Start-Stop-XmasTree is being edited.** The Smart Keyboard supports feedforward.

*Expressive power of the SPOK language*
The language supports a mix of imperative and rule-based programming paradigms. For example, the Start-Stop-XmasTree program shown in sidebar, is comprised of an imperative section composed of actions *Switch off* and *Blink*, and 2 rules. Conditions can be expressed both in terms of states (e.g., *if the blue-lamp is on*) and events (e.g., *each time the blue-lamp is turned on*). Home entities can be denoted by using properties and relations (e.g., *all lamps located in bedroom*). Parallelism is supported at multiple grains: several programs can be started simultaneously; within a program, several rules can fire simultaneously and a program can start/stop the execution of another one.

*Dynamic adaptation*
The grammar of the SPOK language is dynamically updated according to the set of services and devices that are currently available in the home. As a result, the Smart Keyboard, which guides end-users in entering program elements, shows options that are both syntactically valid for the current insertion point, and compliant with the current state of the home (see example in sidebar). Similarly, programs that reference devices and services that are no longer available are flagged so that end-users are made aware that running these programs may result in unexpected behavior. In the example of the Start-Stop-XmasTree shown at the top of the sidebar, the triangle at the top left of the figure indicates that the program is running. The lamp referenced by the *Blink* action has disappeared from the execution environment. Consequently, its reference has been changed to *Unknown* and the triangle has turned from green to orange to indicate that execution continues but at our own risk.

*Debugging aids*
Debugging aids come in three complementary forms: history (by the way of time lines), dependency graph (see sidebar next page), and execution indicators in the source code of SPOK programs. For example, the imperative section of Start-Stop-XmasTree has been executed (counters are set to 1) while no rule has been fired so far (counters are equal to 0) and all of them are waiting for their condition to become true (arrows and light blue background indicate waiting points).

## Lessons from Using SPOK in Our Own Homes

SPOK has been deployed for a period of 4 months in 5 distinct homes of project team members. The intent was to test the robustness and usability of the system while refining the system on a weekly basis. Beyond typical bugs, the main findings are the following: improvements for the preparation of field experiments, discovery of new concepts and key issues for future research, and confirmation of findings about our own behavior consistent with results from the literature. (This last issue will not be developed any further here.)

*Improvements for field experiments*
<u>Harnessing the hardware</u>. Wireless sensors and devices are sensitive to physical conditions (e.g., out of range, lack of power). This sensitivity is generally not "visible". For example, sensors powered with solar cells will fail after a few days without sufficient light. Smart plugs, when installed too far from their dongle, appear and disappear in an unpredictable manner. It is necessary to learn how to build relays between them. The same holds for Hue lamps when blocked by concrete walls. For wireless switches, clicking requires energetic press to generate events. The cover of the DomiCube must be opened to access the led that indicates that it is too

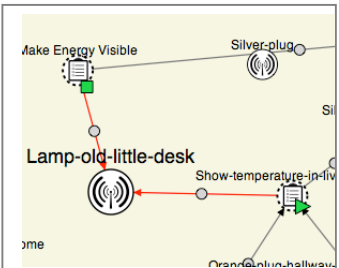

**Extraction of a dependency graph.** The red arrows that converge to *Lamp-old-little-desk* indicate that the programs *Make Energy Visible* and *Show-temperature-in-living-room* both modify the state of the lamp. This can result in unpredictable behavior if the two programs run concurrently. In this example, *Show-temperature-in-living-room* is running (green triangle) while *Make Energy Visible* is not (green square). In this example, *Show-temperature-in-living-room* is running (green triangle) while *Make Energy Visible* is not (green square).

far from its Bluetooth dongle. All these details cannot be discovered in the lab. Work-arounds to such problems must be documented and explained to subjects.

Taming the technology. As St-Exupéry wrote for the fox: "If you tame me, then we shall need each other." Indeed, in the first week after installation, it was common to not find any interesting uses for the system. However the system progressively became an integral part of the home as experience was gained. For example, we have discovered that the hue lamps can provide a tangible representation for temperature and energy consumption. We opportunistically wrote a program to control the lighting of the Christmas tree (as we repeatedly forgot to stop it manually before going to bed). Our conclusion is that the system must be installed in a subject's home for a minimum of one month to report interesting results. In addition, our programs, which emerged from real life needs, can serve as examples for a forum opened to the subjects.

*Discovery for future research and improvement*
Privacy issues made real. By analyzing the time lines provided by SPOK, it was striking to discover how much the rhythm of daily life can be discovered from data recorded by the Trace Manager: movements in the home, arrival and departure, meal and bed times, etc.

Some devices are more critical than others. In the design of SPOK, we missed the notion of "critical" device. This became clear when we used a smart plug to measure the consumption of our refrigerator, while absent for two days. Such a plug must not fail and cut electricity to the refrigerator – as could possibly through program or user. Similarly, some devices require access control (e.g., TV for kids late at night).

## Conclusion
We have two take-away messages. First, longitudinal experiments of a EUDE like SPOK in real-world settings require a middleware that reliably supports dynamic software adaptation and automatic deployment. Secondly, an initial deployment using "our own dog food" in homes of project team members provides highly valuable information for tuning the protocols before the start of field studies, thus improving the quality of the evaluation itself while saving time and discovery for future research.

## Acknowledgements
This work is supported by the European AppsGate CA110 project. We wish to thank our collaborators: R. Barraquand, M. Bidois, S. Caffiau, J.R. Courtois, A. Demeure, J. Estublier, T. Flury, C. Gérard, C. Lenoir, J. Nascimento, K. Pethoukov, C. Roux, G. Vega.